\documentclass{article}
\usepackage{arxiv}
\usepackage{amssymb}
\usepackage{pifont}
\usepackage[utf8]{inputenc} 
\usepackage[T1]{fontenc}    
\usepackage{comment}
\usepackage{hyperref}       
\usepackage{url}            
\usepackage{booktabs}       
\usepackage{amsfonts}       
\usepackage{nicefrac}       
\usepackage{microtype}      
\usepackage{lipsum}		
\usepackage{graphicx}
\usepackage{multirow}
\usepackage{multicol}
\usepackage{todonotes}
\usepackage{natbib}
\usepackage{doi}
\usepackage{subcaption}
\usepackage{amsmath}
\usepackage{booktabs} 

\title{A Survey on 30+ Years of Automatic Singing Assessment and Singing Information Processing}

\date{} 

\author{ \href{https://orcid.org/0000-0002-3989-7105}{\includegraphics[scale=0.06]{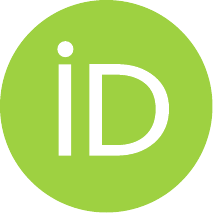}\hspace{1mm}Arthur N. dos Santos}
  \\
	Faculdade de Engenharia Elétrica e de Computação\\
	Universidade Estadual de Campinas\\
	Campinas, Brazil \\
	\texttt{arnics@unicamp.br} \\
	\And
	\href{https://orcid.org/0000-0002-2246-4450}{\includegraphics[scale=0.06]{orcid.pdf}\hspace{1mm}Bruno S. Masiero} \\
	Faculdade de Engenharia Elétrica e de Computação\\
	Universidade Estadual de Campinas\\
	Campinas, Brazil \\
	\texttt{masiero@unicamp.br} \\
}

\renewcommand{\shorttitle}{\raggedright A Survey on 30+ Years of Automatic Singing Assessment and Singing Information Processing}

\hypersetup{
pdftitle={A Survey on 30+ Years of Automatic Singing Assessment and Singing Information Processing},
pdfauthor={Arthur N. dos Santos, Bruno S. Masiero},
pdfkeywords={automatic singing assessment, real-time visual feedback, acoustical biofeedback, singing information processing, karaoke scoring},
}

\begin{document}
\maketitle

\begin{abstract}

Automatic Singing Assessment and Singing Information Processing have evolved over the past three decades to support singing pedagogy, performance analysis, and vocal training. While the first approach objectively evaluates a singer’s performance through computational metrics ranging from real-time visual feedback and acoustical biofeedback to sophisticated pitch tracking and spectral analysis, the latter method compares a predictor vocal signal with a target reference to capture nuanced data embedded in the singing voice. Notable advancements include the development of interactive systems that have significantly improved real-time visual feedback, and the integration of machine learning and deep neural network architectures that enhance the precision of vocal signal processing. This survey critically examines the literature to map the historical evolution of these technologies, while identifying and discussing key gaps. The analysis reveals persistent challenges, such as the lack of standardized evaluation frameworks, difficulties in reliably separating vocal signals from various noise sources, and the underutilization of advanced digital signal processing and artificial intelligence methodologies for capturing artistic expressivity. By detailing these limitations and the corresponding technological advances, this review demonstrates how addressing these issues can bridge the gap between objective computational assessments and subjective human-like evaluations of singing performance, ultimately enhancing both the technical accuracy and pedagogical relevance of automated singing evaluation systems.
\end{abstract}


\section{Introduction}\label{intro}


A singing assessment evaluates a singer’s performance based on both technical and expressive criteria. Traditionally performed by human evaluators, computational approaches now enable objective and accessible evaluation of singing performance. In addition, these methods provide real-time Visual Feedback (VFB) through displays, which guides singers to adjust their performance. Systems that incorporate biological sensors further enhance this process by capturing physiological signals and translating them into visual cues, thereby allowing singers to perceive aspects of their vocal apparatus that are otherwise imperceptible~\cite{Nichols2017, Silas2023}.

Despite these advantages, ASA predominantly relies on metadata-based analysis, such as extracting and comparing pitch values, rather than processing raw vocal signals. These systems neither enhance nor isolate the singer’s voice from the recorded audio, nor facilitate direct comparison with reference vocal tracks, such as original vocals from commercially available songs. Consequently, the emergence of Singing Information Processing (SIP)~\cite{5495212} marks a paradigm shift, treating singing as data by analyzing the actual vocal signal rather than relying solely on the extracted parameters. This transition introduces significant challenges, particularly in voice separation, in which algorithms must accurately isolate a singer’s voice while minimizing artifacts. Errors such as the misclassification of instrumental accompaniment as vocals, overlapping vocal sources, and residual interference from the original track can substantially impact the analytical accuracy~\cite{8683555, araki202530yearssourceseparation}.  


In this survey, a structured framework was adopted to analyze the development of ASA and SIP. The historical analysis focuses on the period 1988–2010. During this time, the seminal contributions by Welch \textit{et al.}~\cite{c32f69d162404332adc463f8106be55a, welch1988singad, 0305735689172005, HOWARD198989, Howard01011997, daw1998real}, Rossiter \textit{et al.}~\cite{howard1993real, rossiter1994albert, 414872, rossiter1995real}, Howard \textit{et al.}~\cite{howard2002quantifying, howard2003towards, Howard01102004, 1321103X050240010401, Howard01102004, welch2004voxed, Welch01072005, HOWARD200720, 1543134}, and Goto \textit{et al.}~\cite{GOTO2004311, 4475948, 5495212} constitute nearly 40\% of the entire body of literature on this topic. Their work is of paramount importance, as it establishes the early methodological and technological foundations of the fields under significant technical constraints. These pioneering studies not only defined the scope of research during that period, but also paved the way for subsequent developments that expanded these initial approaches to a broader range of applications while incorporating contemporary technological advancements.

Subsequently, the remaining literature, spanning up to 2022, is reviewed within a broader scope. Rather than conducting an in-depth historical analysis, this stage focuses on identifying trends by examining key aspects, such as intended applications, feature extraction methods, types of target variables, and assessment methodologies. This shift in approach prioritizes the identification of emerging patterns over their historical significance, thereby facilitating the recognition of potential gaps and future research directions within these fields.  

The remainder of this paper is structured as follows: Section~\ref{sec2} outlines the methodology adopted, including the timeframe, keywords, sources, exclusion criteria, and guiding research questions. Section~\ref{sec3} presents a historical analysis, and Section~\ref{sec4} examines the contemporary state of research in ASA and SIP. Section~\ref{sec5} discusses emerging trends and potential research gaps in these fields and Section~\ref{sec6} provides concluding remarks.

\section{Methodology} \label{sec2}

This survey investigates the development of ASA and SIP systems from their inception to the most recent advancements in the field. The review period spans from late 1980s, corresponding to the earliest documented ASA system, to the early 2020s, with the most recent studies found in this analysis.


A comprehensive literature review was conducted using a keyword-based search strategy in which the following search terms were employed: ``automatic singing assessment'', ``real-time visual feedback'', ``acoustical biofeedback'', ``singing information processing'', and ``karaoke scoring''. These keywords facilitated the retrieval of relevant scholarly works, including journal articles, conference proceedings, and advanced degree dissertations, from databases such as Google Scholar\footnote{\url{https://scholar.google.com/}}, Scopus\footnote{\url{https://www.scopus.com/home.uri}}, ResearchGate\footnote{\url{https://www.researchgate.net/}}, and Academia\footnote{\url{https://www.academia.edu/}}. Although ResearchGate and Academia index some non-peer-reviewed materials, only rigorously peer-reviewed publications were included in the final review.

The literature surveyed spans multiple disciplines, including music education, psychology, acoustics, speech communication, and computational learning. Notable journals include the \textit{Canadian Journal of Music Education}, \textit{Psychology of Music}\footnote{\url{https://us.sagepub.com/en-us/nam/psychology-of-music/journal201640}}, \textit{Music Education Research}\footnote{\url{https://www.tandfonline.com/journals/cmue20}}, \textit{Applied Acoustics}\footnote{\url{https://www.sciencedirect.com/journal/applied-acoustics}}, \textit{The Journal of the Acoustical Society of America}\footnote{\url{https://pubs.aip.org/asa/jasa}}, \textit{Speech Communication}\footnote{\url{https://www.sciencedirect.com/journal/speech-communication}}, and \textit{IEEE Transactions on Audio, Speech, and Language Processing}\footnote{\url{https://ieeexplore.ieee.org/xpl/RecentIssue.jsp?punumber=10376}}.

Additionally, interdisciplinary and computational sources were examined, including the \textit{Journal of Computer Assisted Learning}\footnote{\url{https://onlinelibrary.wiley.com/journal/13652729}}, \textit{Logopedics Phoniatrics Vocology}\footnote{\url{https://www.tandfonline.com/journals/ilog20}}, \textit{Journal of Voice}\footnote{\url{https://voicefoundation.org/}}, and \textit{Journal of Music, Technology \& Education (JMTE)}\footnote{\url{https://www.atmimusic.com/journal-of-music-technology-and-education/}}.

Key conference proceedings relevant to acoustics, speech processing, music technology, and artificial intelligence are also included. These include the \textit{Institute of Acoustics Proceedings}\footnote{\url{https://www.ioa.org.uk/conference-series/institute-acoustics-proceedings}}, \textit{International Conference on Acoustics, Speech, and Signal Processing (ICASSP)}\footnote{\url{https://ieeexplore.ieee.org/xpl/conhome/1000002/all-proceedings}}, \textit{INTERSPEECH}\footnote{\url{https://www.isca-speech.org/}}, \textit{Audio Engineering Society (AES) Convention}\footnote{\url{https://aes2.org/event/conventions/}}, and the \textit{International Society for Music Information Retrieval Conference (ISMIR)}\footnote{\url{https://www.ismir.net/}}.

Furthermore, multidisciplinary conferences such as the \textit{International Symposium on Computer Music Multidisciplinary Research (CMMR)}\footnote{\url{https://link.springer.com/conference/cmmr}}, \textit{Sound and Music Computing Conference (SMC)}\footnote{\url{https://smcnetwork.org/}}, \textit{IEEE Visual Communications and Image Processing (VCIP)}\footnote{\url{https://ieeexplore.ieee.org/xpl/conhome/1800602/all-proceedings}}, and \textit{International Conference on New Interfaces for Musical Expression (NIME)}\footnote{\url{https://nime.org/}} were also reviewed.

This study also considered M.Sc.~and Ph.D. dissertations from institutions known for their contributions to music technology and acoustics, such as the \textit{University of York}\footnote{\url{https://www.york.ac.uk/}} and \textit{The Royal Institute of Technology (KTH)}\footnote{\url{https://www.kth.se/en}}.

Although karaoke scoring systems are relevant to this survey, they are predominantly documented in patents that often lack a theoretical foundation, comprehensive evaluation methodologies, or experimental validation~\cite{tsai2011automatic}. Consequently, patent documentation is excluded from the analysis.

The selected corpus was analyzed based on the following key research questions:

\begin{enumerate}
\item What was the intended application?
\item What feature extraction methods were applied to the predictor signal?
\item Was the target variable represented as data or metadata?
\item How was the assessment conducted (objective or subjective)?
\end{enumerate}  

\section{Historical Analysis}\label{sec3}

Initially, ASA was the primary approach, as capturing a singer’s voice, extracting meaningful features, and comparing them to reference values was already a challenging task, given the technological limitations of the time. However, as technology advanced, SIP emerged, marking a paradigm shift. Unlike ASA, which relies on metadata-based comparisons, SIP compares a vocal predictor signal directly to a vocal target signal, treating singing as data rather than as a derived parameter. This section explores the historical origins of the two intertwined fields.

\subsection{Automatic Singing Assessment (ASA)}

The field of ASA is widely considered to have originated from four foundational systems, each representing the evolution of its predecessor. These systems were developed by closely collaborating with authors and sharing overlapping methodologies to form an interconnected framework. The following subsection describes these pioneering ASA systems in detail.

\subsubsection{SINGing Assessment and Development (SINGAD) system}

Widely regarded as the earliest software for ASA, the SINGing Assessment and Development (SINGAD) system~\cite{c32f69d162404332adc463f8106be55a} was designed to support primary school music education by providing real-time VFB on vocal pitches. It employed a peak-picking device adapted from a cochlear implant, enabling children to visually correlate their fundamental frequency (\(f_0\)) with a target pitch displayed on a monitor~\cite{welch1988singad}.  

Empirical studies~\cite{0305735689172005, HOWARD198989} have indicated that students using SINGAD, even without adult supervision, exhibited greater improvements in singing ability compared to those taught through traditional teacher-student methods. The system was later upgraded to incorporate a Musical Instrument Digital Interface (MIDI) support for audio playback and expanded functionality~\cite{Howard01011997, daw1998real}.  

\subsubsection{Acoustic and Laryngeal
Biofeedback Enhancement in Real-Time (ALBERT)}

While SINGAD was limited to providing real-time VFB on \(f_0\), the Acoustic and Laryngeal Biofeedback Enhancement in Real-Time (ALBERT) software expanded upon its foundation by incorporating a broader range of vocal parameters. It was developed under the hypothesis that effective vocal feedback necessitates the real-time monitoring of additional features, particularly laryngeal activity. Its enhanced capabilities included the display of the laryngeal Closed Quotient (CQ), spectral ratio, Sound Pressure Level (SPL), shimmer, and jitter~\cite{howard1993real, rossiter1994albert, 414872}.  

Experiments conducted as part of a doctoral research project~\cite{rossiter1995real} demonstrated that ALBERT usage led to a general increase in vocal SPL and CQ ratios, which are closely associated with the ``singer’s formant'' and vocal range, respectively. Additionally, ALBERT’s visualizations, optimized through color mapping, proved particularly beneficial for individuals with limited attention spans.  

\subsubsection{Windows SINGAD (WinSINGAD)}

Building upon SINGAD and ALBERT, developers introduced WinSINGAD, a software designed for ASA that was compatible with Microsoft Windows operating systems~\cite{howard2002quantifying, howard2003towards, Howard01102004}. It provided a comprehensive suite of display panels, including the input acoustic pressure waveform, \(f_0\), short-term spectrum, narrow-band spectrogram, spectral ratio, vocal tract area changes, and a sideways webcam feed for visual assessment.  

While WinSINGAD advanced the accessibility of ASA systems, its primary contribution was not technological innovation, but rather its role in facilitating research on singing education~\cite{1321103X050240010401}, particularly in studio environments~\cite{Howard01102004, welch2004voxed, Welch01072005, HOWARD200720} and voice therapy~\cite{1543134}. Studies leveraging WinSINGAD underscored the significance of real-time VFB, with sideways webcam feed proving useful for evaluating posture-related aspects of singing, such as jaw positioning and head/neck alignment.  

However, these studies highlighted several experimental challenges. Excessive reverberation from off-axis microphone placement, signal overload when using \textit{lavalier} microphones in close proximity to the singer’s lips, and acoustic interference from musical accompaniments contributed to measurement errors and potential inaccuracies in the display panels.  

\subsection{Singing Information Processing (SIP)}

At the time of WinSINGAD, other ASA applications such as Sing\&See~\cite{wilson2008learning} also demonstrated the effectiveness of real-time quantitative VFB. However, these systems rely on symbolic representations, such as music scores, or machine-readable metadata, such as MIDI files.  

To overcome these limitations, SIP~\cite{5495212} has emerged to promote the analysis of vocal singing in relation to the vocal components of commercially available CD recordings.

\subsubsection{MiruSinger}

The inauguration of this new field was marked by the development of MiruSinger~\cite{4475948}, a system designed to evaluate \(f_0\) and vibrato characteristics in a user’s singing voice. The system estimated \(f_0\) in real-time by identifying the most dominant harmonic structure in vocal input. For vocal parts in the CD recordings, \(f_0\) estimation was performed using the PreFEst algorithm~\cite{GOTO2004311}, enabling a comparative analysis of the two \(f_0\) trajectories. Vibrato likeliness was quantified by applying a Short-Time Fourier Transform (STFT) to the first-order finite difference of \(f_0\), which served as a key feature. A Support Vector Machine (SVM) classifier was then used to categorize the vibrato quality as either poor or good. In the visual display, the target feature traces were overlaid with user input features, facilitating an assessment of their similarity or divergence.

\section{Contemporary Results}\label{sec4}

Since SINGAD, ALBERT, WinSINGAD, and MiruSinger, numerous studies have expanded upon the ASA and SIP approaches, leading to a diverse range of applications. This section categorizes these contributions based on their primary fields of application and provides a structured overview of contemporary advancements in the domain.

\subsection{Survey Studies}\label{41}

Other survey studies examined the development and pedagogical impact of real-time VFB tools in singing training. 

Hoppe et al.~\cite{hoppe2006development} reviewed four VFB systems — SINGAD, ALBERT, and Sing\&See — analyzing their functional evolution and pedagogical applications. This study focused on their effectiveness in assisting singing and adolescent students by evaluating both quantitative and qualitative measures of training outcomes. While the findings confirmed the benefits of VFB in enhancing singing proficiency, the study emphasized the need for further research to optimize the balance between the level of detail in visual representations, user skill level, and pedagogical integration of these tools.

Expanding on this research, Lã and Fiuza~\cite{la2022real} conducted a comprehensive review of real-time VFB applications, emphasizing the respiratory and oscillatory aspects of voice production. Their analysis encompassed key physiological and acoustic monitoring techniques, including breathing pattern tracking (e.g., lung volume and subglottal pressure), vocal-fold vibratory analysis (via electroglottography and inverse filtering), and acoustic visualization (via spectrographic analysis). Their findings underscored the pedagogical advantages of biofeedback and demonstrated that real-time physiological and acoustic insights contribute to a more scientifically grounded approach to vocal training for both students and educators.

\subsection{Pedagogical Studies}\label{42}

A series of pedagogical studies has evaluated the effectiveness of real-time VFB in singing instruction. These efforts primarily focus on assessing existing technologies rather than advancing new technological frameworks. 

Callaghan \textit{et al.}~\cite{callaghan2001applications} explored the application of computer-assisted VFB in singing instructions and examined its impact on vocal training. Their study targeted both singing students and instructors and analyzed key features such as pitch tracking, vowel quality, and resonance visualization through five distinct display types. The assessments emphasized acoustic analysis, the effectiveness of visual representations, and their pedagogical applicability. The findings underscored the importance of accuracy and task relevance in VFB, highlighting the limitations of speech-optimized algorithms when applied to a broader pitch and intensity range of singing.

Building on this foundation, Callaghan~\cite{callaghan2004science} further investigated the role of VFB in singing instruction by structuring the analysis of three key aspects: (a) extraction of perceptually relevant vocal characteristics (e.g., pitch, vowel identity, and timbre), (b) development of meaningful visual displays to aid singers in adjusting their vocal outputs, and (c) pedagogical integration of VFB into structured singing lessons. The study employed commercially available Kay Elemetrics\footnote{\url{https://www.kayelemetrics.com/}} tools, such as Visi-Pitch for pitch tracking, Sona-Match for vowel identity, and Computerized Speech Lab (CSL) for timbre analysis, all installed on a dedicated system. To ensure consistency, a high-quality headset microphone was used to prevent postural adjustments that might affect the fixed microphone setup to ensure consistency. These findings contribute to a deeper understanding of how real-time VFB can be effectively leveraged to support vocal development and pedagogy.

Wilson \textit{et al.}~\cite{wilson2008learning} examined the impact of real-time VFB on improving singing pitch accuracy by comparing two feedback modalities: a grid-based system that provides target-based visual guidance and a keyboard-style display that indicates real-time pitch without historical tracking. These methods were evaluated against a control group that did not receive feedback. Pitch accuracy, measured in cents, showed significant improvements in both experimental groups, whereas the control group showed no measurable progress. This study demonstrated the critical role of augmented feedback in skill acquisition, reinforcing the necessity of visual support for refining pitch accuracy in singing.

Martínez \textit{et al.}~\cite{martinez2019analysis} conducted an objective analysis of vocalizations for singing pedagogy, focusing on \(f_0\) as the primary feature. Their study analyzed sustained vowel phonations across three temporal zones — initial, middle, and final — to assess the vibrato characteristics and unintended oscillations. The evaluation was performed using Praat\footnote{\url{https://www.fon.hum.uva.nl/praat/}} for \(f_0\) extraction and MATLAB\footnote{\url{https://www.mathworks.com/products/matlab.html}} for visualization. The results indicated that the combined use of auditory stimuli and VFB significantly enhanced vocalization interpretation, suggesting that real-time visual reinforcement fosters more effective learning than auditory perception alone.

\subsection{Singing Practice}\label{43}

Beyond pedagogical applications, real-time VFB has been extensively utilized in singing practice with a focus on improving vocal performance through acoustic analysis, pitch tracking, and interactive training tools. 

Lundy \textit{et al.}~\cite{lundy2000acoustic} conducted an acoustic analysis of singing and speaking voices among singing students, employing the Singing Power Ratio (SPR) as a primary metric to differentiate between them. The study involved singers at various training levels, with digital recordings analyzed using Fast Fourier Transform (FFT) and Multi-Dimensional Voice Program (MDVP) software. The extracted features included SPR, \(f_0\), jitter, shimmer, and Noise-to-Harmonic Ratio (NHR). Contrary to prior findings on professional singers, the results revealed no significant difference in SPR between sung and spoken tones. However, speech samples exhibited higher shimmer and NHR values, suggesting greater spectral variability in the spoken voice.

The Singing Coach system~\cite{randall2006singing} was developed to assist singers in improving pitch and tempo through real-time pitch tracking and visualization. Designed for users across beginner, intermediate, and advanced levels, this tool incorporates pitch tracking, visual notation, metronome guidance, and playback functionalities for self-assessment. Evaluations were conducted objectively with the system providing automatic scoring and immediate feedback. Training difficulty progressively increased to promote greater precision in vocal performance.

Lin \textit{et al.}~\cite{lin2014implementation} introduced a real-time interactive pitch training tool for iOS aimed at enhancing singers' intonation and timing. The system employs FFT and \textit{cepstrum}-based pitch estimation, incorporating an intonation classifier, scoring mechanism, and interactive training module. Targeting individuals seeking to refine pitch accuracy, the assessment was performed using real-time VFB and a pass/fail scoring system based on pitch deviation within a predefined threshold. The objective evaluation demonstrated significant user improvement, with an average intonation accuracy of 94.81\% following the training.

Cantus~\cite{cantus}, a web-based application for real-time vocal training, utilizes a listen-and-repeat approach with a synchronized VFB. Users sang into a microphone while observing the illuminated notes on the screen, aligning their vocal output with the reference displayed. The pitch accuracy assessment relies on the YIN algorithm~\cite{yin} for \(f_0\) detection by comparing the user input with the target notes. The scoring system ranged from 0 to 10, with a maximum score awarded for pitch deviations within one-third of the semitone. The evaluation method was objective, leveraging frequency analysis to quantify the performance accuracy.

Rosenzweig \textit{et al.}~\cite{rosenzweig2020tunein} developed TuneIn, a web-based interface designed to support choir rehearsals using real-time intonation feedback. The system analyzed pitch accuracy using \(f_0\) estimation and deviation analysis, specifically targeting choral performance accompanied by piano. The assessment process employed the Convolutional Representation for Pitch Estimation (CREPE) algorithm~\cite{kim2018crepeconvolutionalrepresentationpitch} for real-time pitch estimation and visualization of deviations within a piano roll interface. Objective evaluations measured intonation deviation in cents and provided color-coded feedback to assist singers in refining their pitch accuracy.

\subsection{Games and Virtual Reality}\label{44}

Interactive game-like systems and Virtual Reality (VR) applications have also been explored as innovative tools for vocal training, offering immersive and interactive environments that enhance user engagement and performance assessments. 

Brereton \textit{et al.}~\cite{brereton2014singing} developed the Virtual Singing Studio (VSS), which is a system designed to simulate different acoustic environments and analyze how singers adjust their performance to varying spatial conditions. The system provided real-time aural feedback, enabling singers to perceive their own voices as if they were performing live. The study involved professional singers performing in both real and virtual spaces, with assessments conducted through objective signal-processing techniques, measuring tempo, vibrato, and intonation, and subjective evaluations based on user feedback. While singers reported noticeable differences between real and virtual environments, objective results indicated limited performance variations, with the vibrato extent being the most affected parameter.

Al Kork \textit{et al.}~\cite{al2015novel} introduced an interactive game-like system aimed at teaching, performing, and evaluating human beatboxing. The system leveraged real-time sensor data to analyze tongue movements, facial gestures, and vocal pitch, thereby providing an engaging and structured learning environment. A helmet-basd setup, integrating an ultrasonic transducer, a Kinect camera, and a microphone, was used to track articulatory movements and compare it with expert models. The assessment framework combined objective measurements, such as motor behavior and pitch accuracy, with subjective feedback from a virtual tutor who delivered oral and written guidance to refine user performance.

Zhang \textit{et al.}~\cite{zhang2021real} proposed a real-time singing scoring system utilizing VR, \(f_0\) analysis, and Dynamic Time Warping (DTW) for audio comparison. The system was designed for singers practicing in a VR setting to enhance their stage presence and alleviate their performance anxiety. Assessments were conducted by analyzing users' vocal performances against reference tracks, with real-time scores influencing character animations on a virtual stage. The objective evaluation cohigh scoring approach, while subjective feedback highlighted strong immersion and the effectiveness of the system in reducing anxiety about virtual performance.

\subsection{Karaoke Scoring}\label{45}

Despite the relative scarcity of scientific literature on karaoke scoring, few notable studies have explored methods for automatic evaluation, focusing on statistical modeling, pitch accuracy, rhythm, and dynamic analysis.

Qiu~\cite{Qiu538862} developed a karaoke scoring algorithm employing Hidden Markov Models (HMM) trained with Gaussian Mixture Models (GMM) to evaluate musical features such as pitch, accent, zero-crossing rate, and power. The system utilizes the YIN algorithm~\cite{yin} for pitch detection and introduces a semitone tolerance parameter to adjust the difficulty levels. The training was based on a dataset recorded by an amateur yet well-trained singer, with intra-note models developed using the HTK toolbox\footnote{\url{https://htk.eng.cam.ac.uk/}}. During the evaluation, the system segmented the singing input into individual notes and scored the accuracy based on an error rate, defined as the ratio of misclassified frames to total frames. This approach relies on statistical modeling to provide an objective performance assessment.

Tsai \textit{et al.}~\cite{tsai2011automatic} proposed an automatic singing evaluation system for karaoke performance, integrating pitch, volume, and rhythm features extracted from karaoke VCD recordings. The system employed a multi-tiered scoring methodology: pitch-based ratings were determined using MIDI notes and DTW; volume-based ratings were derived from log-energy sequences analyzed with DTW; and rhythm-based ratings were computed using HMMs. The objective evaluation demonstrated a strong correlation (Pearson coefficient of 0.82) between the system's automatic scores and human ratings, validating its effectiveness in performance assessment.

Narang \textit{et al.}~\cite{narang2022analysis} focused on the automatic transcription of dynamic markings in vocal rock and pop performances, using loudness as the primary evaluation feature. This study analyzed commercial vocal recordings and their corresponding karaoke versions by extracting source-separated vocal tracks and comparing them with the original vocal stems. An assessment was conducted using the Pearson Correlation Coefficient to measure the similarity between loudness curves, supplemented by peak-based histogram analysis for local dynamics and global dynamic range computation. The evaluations incorporated both objective (single-scale loudness metrics) and subjective (expert assessments of informativeness and accuracy) criteria. Although the approach successfully captured structural similarities in dynamics, artifacts from source separation processes introduced limitations in accuracy.

\subsection{Talent Shows and Conservatory Entrance Assessments}\label{46}

Despite serving distinct purposes, talent shows and conservatory entrance assessments share a common methodology in which candidates' singing abilities are evaluated by an examination board. In this context, some studies have explored automated alternatives to human evaluation.

Lal~\cite{lal2006comparison} developed a telephone-based singing evaluation system designed for televised talent shows. The system compares users' renditions of pre-recorded song clips against their original versions, employing the Robust Algorithm for Pitch Tracking (RAPT)~\cite{talkin1995robust} for pitch estimation and Viterbi alignment to measure pitch differences. These differences were quantified using the Euclidean distance metric, which was subsequently transformed into a rating on a ten-point scale. Users were assessed based on pitch curve alignment, with substitution errors penalized accordingly. Both objective (algorithmic scoring) and subjective (human judge rankings) evaluations were conducted, revealing high engagement as a significant proportion of calls originated from repeat participants.

Bokshi \textit{et al.}~\cite{bokshi2017assessment} introduced an automatic singing quality classification system to evaluate contestants in televised American singing competitions. The system utilizes both audio and visual features, with Mel-Frequency Cepstral Coefficients (MFCCs) representing the audio component and lip/eye movement analysis to form the visual component. Candidates comprising both trained and untrained singers were classified as either qualified or non-qualified based on judgment. The assessment was performed using machine learning classifiers, including Logistic Regression, Naïve Bayes, and k-Nearest Neighbors (k-NN), applied to unimodal (audio, lip, or eye) and multimodal (audio-visual) features. The results indicated that audio-based classification achieved 95\% accuracy, whereas the inclusion of lip and eye features modestly improved performance by up to 2\%, with these visual modalities achieving an independent accuracy of 82\%.

Bozkurt \textit{et al.}~\cite{bozkurt2017dataset} proposed a singing voice assessment system designed for conservatory entrance exams, utilizing \(f_0\) analysis as the primary evaluation feature. The system extracted \(f_0\) series from a dataset comprising 1018 singing recordings and 2599 piano recordings, employing the Melodia algorithm\footnote{\url{https://www.upf.edu/web/mtg/melodia}} for pitch tracking. The performance assessment was conducted using DTW and a machine-learning-based classification system implemented in Julia. The objective evaluation involved training a classifier on DTW-derived distance histograms, achieving an average cross-validation accuracy of 0.74. Owing to privacy constraints, only the extracted \(f_0\) series data were shared rather than the original audio recordings.

\subsection{Rare Singing}\label{47}

Rare singing encompasses vocal traditions, styles, or repertoires that are uncommon, specialized, or preserved within specific cultural, historical, or ritual contexts. Some studies have explored methods for analyzing, teaching, and preserving these unique singing practices using computational tools and interactive systems.

Elmer and Elmer~\cite{elmer2000new} introduced a computer-aided method to analyze and represent unconventional singing styles using pitch and time as primary analytical features. The study focused on sung performances assessed through spectral analysis and \(f_0\) extraction, employing a custom pitch analysis program integrated with CoolEdit\footnote{\url{https://www.adobe.com/special/products/audition/syntrillium.html}}. Objective evaluation combines acoustic and auditory analyses, enabling the visualization of pitch patterns over time while capturing qualities such as pitch stability and instability. The system also facilitated the representation of pitch organization, glissandi, breathing patterns, joint singing, and instructional elements using a novel symbolic notation system, thereby providing a structured approach for nonstandard vocal techniques.

Stavropoulou \textit{et al.}~\cite{stavropoulou2014effectiveness} investigated the effectiveness of VFB-based singing software in improving pitch accuracy and vocal quality among Greek elementary school children (ages 6–9). Using Singing Coach Pro as a training tool, the study assessed participants’ singing abilities by recording performances of both a traditional Greek song and a Western classical piece and analyzing pitch accuracy before and after training. The participants were classified into three groups: those who initially sang well and showed minimal improvement, those who exhibited moderate improvement, and those with significant pitch deviations. The results demonstrated that visual feedback, particularly when combined with teacher guidance, enhanced motivation and singing accuracy, with more pronounced improvements observed for the traditional Greek song and among older children, especially girls.

Chawah \textit{et al.}~\cite{chawah2014educational} developed an educational platform designed to facilitate the learning and analysis of rare singing techniques by capturing multimodal articulatory and acoustic features through a specialized ``hyperhelmet''. The system targeted professional singers performing Corsican \textit{Cantu in Paghjella} and Byzantine chants, with assessments based on synchronized recordings of vocal fold motion, respiration, tongue and lip movements, and audio signals. Data processing was conducted using RTMaps\footnote{\url{https://intempora.com/products/rtmaps/}} and MATLAB-based tools. While the project remained in the data collection phase, its long-term objective was to support the preservation of rare vocal traditions by providing auto-adaptive singing modules and 3D vocal tract visualization for enhanced learning.

Moschos \textit{et al.}~\cite{moschos2016fonaskein} introduced FONASKEIN, an interactive software application for real-time singing voice practice with VFB, particularly at the micro- and non-tempered scales. Developed in Max/MSP, the system enabled users to study microtonal tuning systems, such as Byzantine and Ancient Greek scales, while also allowing the import of custom scales defined in cents. The platform comprises two core components: (1) real-time analysis and transformation of microphone input using the fiddle\~ Max object for pitch detection, and (2) the conversion of MIDI files into musical scores with microtonal playback control. Pitch estimation was performed via a Discrete Fourier Transform (DFT)-based method using the maximum-likelihood approach to extract \(f_0\). By providing real-time auditory and VFB intonation, FONASKEIN supported singers in mastering microtonal tuning systems and refining their pitch accuracy.

\subsection{Acoustical Biofeedback}\label{48}

Building on the foundations established by ALBERT and WinSINGAD, a study explored the use of acoustical biofeedback in singing pedagogy. This approach typically involves the use of wearable devices or intrusive sensors to monitor physiological and acoustic parameters and provide real-time feedback to enhance vocal training.

Piao and Xia~\cite{piao2022sensing} proposed a multimodal singing tutoring system that integrates a wearable breath detector equipped with pressure sensors to monitor breathing states alongside pitch analysis. Designed for elementary singing learners (ages 18–26), the system visually displays pitch contours and breath states on an interactive score interface, allowing real-time feedback during practice. The study employed a 2×2 factorial design with 14 participants (ten with musical backgrounds and four without) who learned two songs under two conditions: pitch-only and pitch + breath feedback. An objective evaluation indicated a 21.25\% improvement in pitch accuracy among musically trained participants when breath monitoring was included, whereas those without formal musical training showed no significant improvement.

\subsection{Expressivity and Expressiveness}\label{49}

Although often used interchangeably, expressivity represents distinct aspects of vocal performance. Expressivity pertains to the temporal dynamics of a performance, encompassing parameters such as attack, decay, sustain, and release (ADSR) envelopes. In contrast, expressiveness is associated with the emotional and interpretative qualities of a performance, which, according to the Karaoke World Championship (KWC) guidelines\footnote{\url{https://www.kwc.fi/}}, involves conveying emotion and passion through vocal delivery, body language, and facial expressions.

Bonada \textit{et al.}~\cite{bonada2006singing} introduced \textit{A Singing Tutor}, an automatic evaluation tool for assessing tuning, tempo, and expression at both note and intra-note levels. The system employs an untrained HMM-based segmentation algorithm supplemented by heuristic rules to align the performance metrics with a MIDI reference. It further identifies expressive elements, such as vibrato, attack characteristics, and note transitions. The system was evaluated using the performances of three commercial pop songs sung by amateur singers, analyzing over 1000 notes. The segmentation achieved 95\% accuracy compared with manual annotations (with a tolerance window of 30\,ms). The expression transcription was manually reviewed by an expert, which yielded an accuracy of over 95\% for correctly segmented notes.

Yu \textit{et al.}~\cite{yu2015performance} proposed a singing voice performance scoring system that integrates acoustic and emotional features extracted using openSMILE\footnote{\url{https://www.audeering.com/research/opensmile/}}. The system evaluates amateur performance by comparing it against professional renditions, employing DTW for alignment, followed by feature extraction (e.g., pitch, loudness, and MFCCs). Classification was performed using SVMs and Deep Boltzmann Machines (DBM). The DBM model demonstrated a slight improvement over the SVM classifier, achieving the highest weighted recall of 0.5854 versus 0.5444. However, the study noted that the overall performance of both models was constrained by the limited dataset size.

\subsection{Other ASA and SIP}\label{410}

Other studies have focused on the development of ASA and SIP systems by leveraging machine learning and signal processing techniques to evaluate vocal performance. These approaches aim to objectively quantify singing quality by analyzing pitch accuracy, note transitions, and overall musical expressiveness.

Zhang \textit{et al.}~\cite{zhang2021learn} proposed a deep metric learning-based ASA system that utilizes a triplet network to map high-quality singing performances closer to a reference track while distinguishing poor performances. The system processes spectrograms, chroma features, \(f_0\), and tempograms using a Convolutional Neural Network (CNN) with self-attention mechanisms. The model was trained and tested on solo singing clips, primarily from Chinese vocal performance. The assessment was conducted by comparing embeddings of high- and low-quality singing relative to a reference track, demonstrating a strong correlation with human ratings. The proposed system outperformed conventional rule-based methods in predicting the singing quality, with objective evaluations confirming the robustness of the learned embedding space.

Faghih \textit{et al.}~\cite{faghih2022new} introduced a novel algorithm for detecting note onsets, offsets, and transitions in solo singing designed for both real-time and offline applications. Their method relies on pitch contour trajectory changes to identify the note boundaries. The system was evaluated using annotated vocal datasets (Erkomaishvili\footnote{\url{{https://www.audiolabs-erlangen.de/resources/MIR/2019-GeorgianMusic-Erkomaishvili}}} and SVNote1\footnote{\url{https://zenodo.org/records/7061507}}) and benchmarked against eight existing onset detection algorithms through F-measure scores and statistical analysis (ANOVA). The proposed algorithm achieved superior performance, outperforming the competing methods by margins ranging from 2\% to 36\%, with the most significant improvements observed in high-quality audio recordings. Additionally, real-time evaluations demonstrated processing delays between 46 and 230\,ms depending on the dataset characteristics.

\section{Discussion}\label{sec5}

This discussion serves two primary objectives: (1) to synthesize the strengths and key contributions of the existing literature and (2) to identify potential research gaps that warrant further investigation.

\subsection{Summary of the Existing Literature}

The studies reviewed in Section~\ref{41} traced the evolution of VFB tools and their integration into contemporary singing pedagogy. They underscored the importance of refining feedback mechanisms to enhance pedagogical effectiveness, ultimately improving both the technical and expressive dimensions of vocal training.

Section~\ref{42} substantiates the pedagogical value of VFB in vocal instruction. These studies emphasize the necessity of perceptually meaningful visual representations and highlight the role of technological advancements in optimizing VFB systems for more effective training methodologies.

The studies in Section~\ref{43} demonstrate the efficacy of VFB-based singing practice tools in improving pitch accuracy, timing, and vocal precision. Their findings reinforced the significance of real-time feedback in structuring vocal training, benefiting both individual performers and ensemble settings.

Section~\ref{44} explores the integration of VR and gamification into vocal training, illustrating their capacities to provide immersive feedback, refine vocal techniques, and enhance learner engagement. By merging real-time analysis with interactive components, these systems represent promising avenues for digital vocal pedagogy.

The studies reviewed in Section~\ref{45} examined diverse methodologies for karaoke scoring, ranging from statistical modeling to feature-based performance evaluation. These contributions pave the way for more sophisticated and reliable ASA systems with applications in music education, gaming, and entertainment.

Section~\ref{46} discusses the advancements in automated singing evaluations, particularly in talent competitions and conservatory admissions. By incorporating pitch tracking, statistical modeling, and Machine Learning (ML)-based classification, these methods offer objective and scalable alternatives to traditional human assessments, thereby improving the consistency and efficiency of vocal performance evaluations.

The studies in Section~\ref{47} focused on the computational analysis, training, and preservation of rare singing traditions. Utilizing pitch tracking, multimodal analysis, and interactive learning tools, these studies demonstrate the potential of technology in both the study and transmission of nonstandard vocal techniques, ensuring their continued cultural relevance.

Section~\ref{48} highlights the role of acoustic biofeedback systems in vocal training, showing their effectiveness in improving pitch accuracy, breathing control, and overall pedagogical outcomes. Through real-time physiological monitoring and interactive feedback mechanisms, these systems provide valuable insights into vocal techniques and performance optimization.

The studies in Section~\ref{49} advance the computational analysis of vocal expressivity by developing methodologies for evaluating the performance dynamics, emotional content, and expressive nuances in singing. These contributions support the development of refined data-driven assessment tools for vocal training and performance evaluation.

Finally, Section~\ref{410} presents advancements in automatic singing evaluation, focusing on improved note-segmentation accuracy and enhanced ML models for performance assessment. These developments have contributed to the development of precise and efficient computational tools for vocal training and music education.

\subsection{Potential Research Gaps for Future Contributions}

The existing literature underscores the increasing role of technology in vocal training, performance assessment, and preservation of diverse singing traditions. As digital tools continue to evolve, future research should focus on refining these methodologies, enhancing accessibility, and integrating them into mainstream pedagogical frameworks. While substantial progress has been made, modern advances in Digital Signal Processing (DSP) and Artificial Intelligence (AI) remain underutilized within the ASA and SIP frameworks. The integration of these technologies has the potential to drive significant advancements in this field.

\begin{enumerate}
    \item \textbf{Predictor and Target Signal Separation}: 
    \\ \\
    A fundamental challenge within ASA and SIP involves mitigating various forms of noise that interfere with vocal signal analysis. These include:
 
    \begin{itemize}   
    \item[a)] \textit{Convolutive Noise}: In real-world environments, sound waves reflect off surfaces such as walls and furniture, introducing acoustic phenomena such as echoes and reverberation~\cite{filipanits1994design, da2003acustica}. These effects distort the recorded signal because of the interaction between the direct sound and reflected components~\cite{allen1979image, oppenheim1997signals}.

    \item[b)] \textit{Additive Noise}: Singing performances often occur in settings with background music, typically played through a Public Address (PA) system or an accompanying musical instrument. Even when using directional microphones, such as cardioid-pattern dynamic microphones~\cite{zotter2019ambisonics}, a certain degree of background noise is inevitably captured by introducing undesired components into the recording.

    \item[c)] \textit{Competitive Noise}: The presence of an audience can introduce further interference in the form of cheering, clapping, or even singing along. This constitutes a particularly challenging form of additive noise as it shares spectral and temporal characteristics with the primary vocal signal, complicating the separation process.
    \end{itemize}

    These noise sources introduce unwanted artifacts into the predictor signal, which ideally represents the singer’s isolated vocal performance. Consequently, advanced enhancement or separation algorithms are required to filter out noise and preserve the integrity of vocal recordings.
    \\ \\
    A similar challenge pertains to the target signal in the SIP, where the goal is to extract isolated vocals from commercially produced songs for comparison with the predictor. This necessitates the development of robust source-separation algorithms capable of effectively isolating vocals from a given music track.
    \\ \\
    Traditional signal separation techniques, such as Wiener filtering, beamforming, and time-frequency masking, have been employed in this context. However, recent advances in data-driven methodologies, particularly AI-driven approaches, have led to significant improvements. The emerging field of ``Machine Hearing''~\cite{lyon2017human}, analogous to Computer Vision, leverages ML and Deep Learning (DL) to achieve high-level auditory scene analysis.
    \\ \\
    One notable approach is Wave-U-Net~\cite{stoller2018wave}, an architecture inspired by Temporal Convolutional Networks (TCNs)~\cite{bai2018empirical} that performs end-to-end source separation directly in the time domain. Configured as a Single-Input Multiple-Output (SIMO) system, it can separate distinct musical components, such as vocals, drums, bass, and other instruments.
    \\ \\
    More recent methods employ 2D deep U-Net architectures~\cite{city19289} for vocal and instrumental separation, utilizing STFT-derived magnitude spectrograms. These models apply encoder-decoder structures with a bottleneck layer to predict spectrogram masks for vocal and instrumental components by using $\ell_1$ loss functions. The continued refinement of these architectures, particularly in terms of generalization across diverse musical styles and recording conditions, remains an open research challenge.
    
    \item \textbf{Features, Computational Metrics, and Non-Intrusive Subjective Evaluation}: 
    \\ \\
    An effective and standardized evaluation framework for singing assessment should integrate technical proficiency and artistic expressivity. Current computational analyses have predominantly focused on low-level features related to technical execution, including pitch accuracy, rhythmic precision, dynamics, breathing control, and vibrato, all of which contribute to high-level musical interpretations. Typical computational descriptors include $f_0$ trajectory analysis, DTW, $\ell_2$ distance metrics, and HMM-based segmentation.
    \\ \\
    However, a more comprehensive evaluation strategy should incorporate a broader spectrum of low-level features that are commonly used in Music Information Retrieval (MIR). These include harmonic representations, such as chromagrams, which capture high-level harmonic relationships, as well as auditory-based spectral representations, including the Bark, Mel, and Gammatone spectrograms. Additionally, their corresponding \textit{cepstral} variants — Bark Frequency \textit{Cepstral} Coefficients (BFCC), Mel-Frequency \textit{Cepstral} Coefficients (MFCC), and Gammatone \textit{Cepstral} Coefficients (GTCC) — offer refined acoustic descriptors that can further enhance the robustness of computational singing assessments.
    \\ \\
    To establish a rigorous and standardized evaluation protocol, qualitative studies should systematically assess the applicability of these computational features across a well-curated test set. This ensures consistency in performance evaluation across different assessment paradigms. Furthermore, artistic expressivity encompassing factors such as emotional conveyance, timbral variation, and lyrical articulation should be integrated into computational assessment models.
    \\ \\
    Advanced Artificial Neural Network (ANN)-based methodologies provide promising avenues for non-intrusive subjective evaluations. For instance, Automatic Speech Recognition (ASR)~\cite{wav2vec} can be employed to assess lyrical precision, whereas Song Emotion Recognition (SER)~\cite{Benevenuto_2022} models can quantify emotional expressivity. Given the state-of-the-art classification performance of contemporary ANNs, incorporating these methods into ASA and SIP frameworks would significantly enhance the depth and reliability of vocal performance assessments.

\end{enumerate}

\section{Conclusion}\label{sec6}

Real-time VFB constitutes a fundamental component of both the ASA and SIP frameworks. While scoring mechanisms, such as those employed in karaoke systems, provide singers with an overall assessment of their performance, immediate feedback on mistakes is crucial to facilitate real-time self-correction and skill development. Despite the demonstrated effectiveness of these approaches in domains such as pedagogy and vocal training, significant advancements are still needed to leverage the recent technological progress in DSP and AI. One of the most pressing challenges is the enhancement and separation of vocal signals, as real-world recording environments introduce various types of noise that can hinder the accuracy and effectiveness of ASA and SIP. Although computational feature extraction in MIR has played a crucial role in vocal analysis, existing studies have only begun to explore a vast range of available descriptors. Further research is required to identify the most suitable features for robust and reliable vocal assessment. Moreover, AI-driven methodologies offer substantial potential for complementing traditional evaluation approaches by modeling the subjective aspects of singing performance, such as lyrical precision and expressive delivery. The integration of AI-based techniques can bridge the gap between computational assessments and human-like evaluations, thereby enhancing the comprehensiveness and accuracy of automated singing analysis systems.

\section{Acknowledgments}
Sincere gratitude is extended to Eng. Mateus Prauchner Enriconi, M. Sc., for his insightful contributions, constructive discussions, and valuable suggestions. 

\section{Data Availability Statement}

The data that support the findings of this study are available from the corresponding author, dos Santos, A. N., upon reasonable request.

\bibliographystyle{unsrt}
\bibliography{references} 
\end{document}